\documentclass[conference]{IEEEtran}
\usepackage{amssymb}
\usepackage{color}
\usepackage{multirow}
\usepackage{mathrsfs}
\newtheorem{theorem}{Theorem}

\newtheorem{definition}{Definition}

\usepackage{changebar}
\usepackage{cite}
\ifCLASSINFOpdf
  \usepackage[pdftex]{graphicx}
  \graphicspath{{../pdf/}{../jpeg/}}
  \DeclareGraphicsExtensions{.pdf,.jpeg,.png}
\else
  \usepackage[dvips]{graphicx}
  \graphicspath{{../eps/}}
  \DeclareGraphicsExtensions{.eps}
\fi
\usepackage[cmex10]{amsmath}
\usepackage{array}

\usepackage[tight,footnotesize]{subfigure}
\usepackage{stfloats}

\usepackage{arydshln}
\usepackage{rotating}
\usepackage{microtype}
\def\gap{.4ex}
\abovedisplayskip\gap
\belowdisplayskip\gap
\abovedisplayshortskip\gap
\belowdisplayshortskip\gap

\IEEEoverridecommandlockouts

\usepackage{epstopdf}
\begin{document}
\title{A Unified Scheme for Two-Receiver Broadcast Channels with Receiver Message Side Information}

\author{\IEEEauthorblockN{Behzad Asadi, Lawrence Ong, and Sarah J.\ Johnson}\thanks{This work is supported by the Australian Research Council under grants FT110100195, FT140100219, and DP150100903.}
	\IEEEauthorblockA{School of Electrical Engineering and Computer Science, The University of Newcastle, Newcastle, Australia}
	Email:{ behzad.asadi@uon.edu.au, lawrence.ong@cantab.net, sarah.johnson@newcastle.edu.au}
}
\maketitle

\begin{abstract}
This paper investigates the capacity regions of two-receiver broadcast channels where each receiver (i) has  both common and private-message requests, and (ii) knows part of the private message requested by the other receiver as side information. We first propose a transmission scheme and derive an inner bound for the two-receiver memoryless broadcast channel. We next prove that this inner bound is tight for the deterministic channel and the more capable channel, thereby establishing their capacity regions. We show that this inner bound is also tight for all classes of two-receiver broadcast channels whose capacity regions were known prior to this work. Our proposed scheme is consequently a unified capacity-achieving scheme for these classes of broadcast channels.
\end{abstract}
\vspace{2pt}
\vspace{-12pt}
\IEEEpeerreviewmaketitle
\section{Introduction}
We investigate the capacity regions of two-receiver broadcast channels~\cite{BC} with receiver message side information where each receiver may know some of the transmitted messages a priori. These channels are of interest due to applications such as multimedia broadcasting with packet loss, and the downlink phase of two-way relay channels~\cite{TWRC}. The capacity regions of these channels are known for only the following special classes of the two-receiver \textit{memoryless} broadcast channel.
\begin{enumerate}
\item Specific message request and side information configuration (for all types of the channel):
\begin{enumerate}
\item Complementary side information: both receivers need to decode all the source messages, i.e., all the messages not known a priori~\cite{BCwithSI2UsersOechtering,SWoverBC}
\item Degraded message sets: one receiver needs to decode all the source messages, and the other one only a subset of the source messages~\cite{BCwithSI2UsersKramer}
\end{enumerate}
\item Specific channel type (for all possible message requests and side information configurations):
\begin{enumerate}
	\item Additive white Gaussian noise (AWGN) channel~\cite{BCwithSI2UsersGeneral}
	\item Less noisy channel
\end{enumerate}
\end{enumerate}
The capacity region for the less noisy case is obtained from the capacity region of the three-receiver less noisy broadcast channel where (i) only two receivers possess side information, and (ii) the request of the third receiver is only restricted to a common message demanded by all the receivers~\cite[Theorem~3]{BCwithSI3UsersCommonMessage}.

The other results obtained from the existing capacity results for broadcast channels with three or more receivers,~\cite{SWoverBC,BCwithSI3UsersCommonMessage,BCwithSI3UsersLinearDeterministic,Capacity3UsersPrivateMessage,Group4andGroup7,Capacity3UsersPrivateMessageParallel}, fall into the mentioned results for two-receiver broadcast channels with complementary side information or degraded message sets.

\begin{figure}[t]
	\centering
	\hskip-2pt\includegraphics[width=0.49\textwidth]{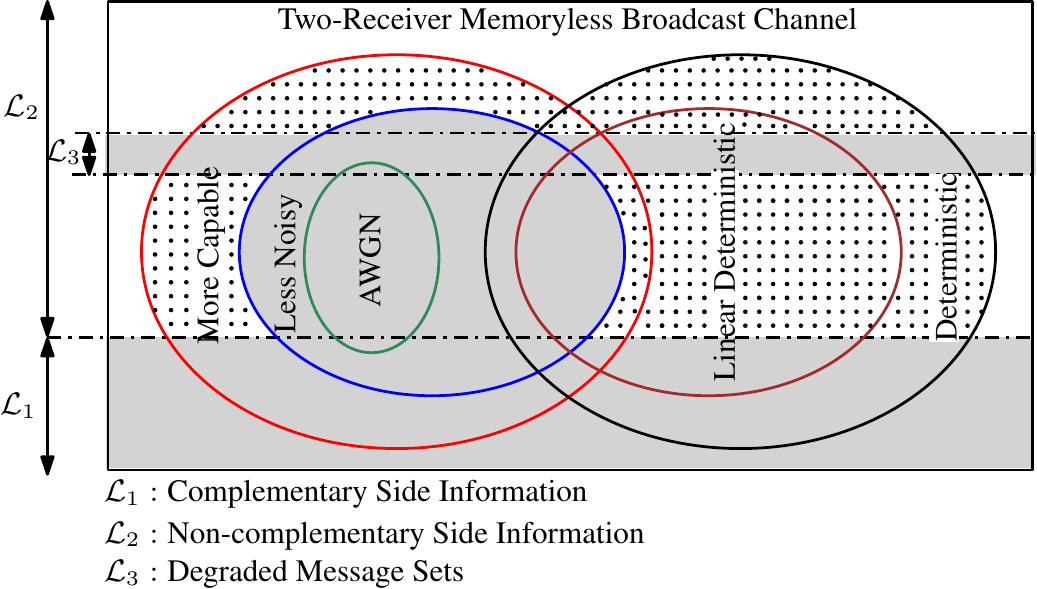}
	\vspace{-20pt}
	\caption{Summary of the results where grey areas represent classes whose capacity regions were known prior to this work, and dotted areas represent classes whose capacity regions are established in this work for all possible message requests and side information configurations (the rectangle represents the two-receiver memoryless broadcast channel, and each oval represents one type of channel).} 
	\vspace{-15pt}
	\label{Fig:ResultSummary}
\end{figure}
\begin{figure}[b]
	\centering
	\vspace{-17pt}
	\includegraphics[width=0.49\textwidth]{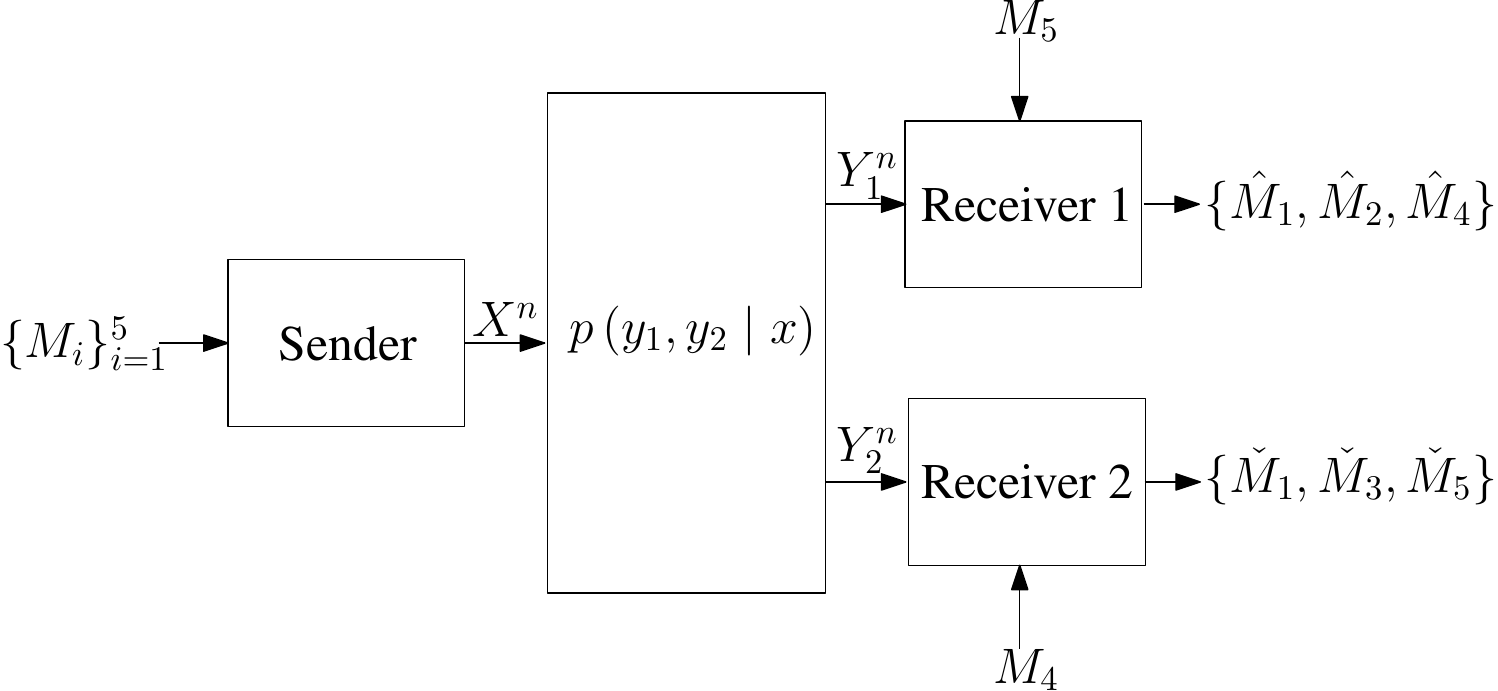}
	\vspace{-20pt}
	\caption{The two-receiver memoryless broadcast channel $p(y_1,y_2\mid x)$ with receiver message side information. $\mathbf{M}=\{M_i\}_{i=1}^5$ is the set of independent messages transmitted by the sender. $\mathbf{K}_1=\{M_5\}$ and $\mathbf{K}_2=\{M_4\}$ are the set of messages known a priori to receivers 1 and 2 respectively. $\mathbf{W}_1=\{M_1,M_2,M_4\}$ and $\mathbf{W}_2=\{M_1,M_3,M_5\}$ are the set of messages requested by receivers 1 and 2 respectively. $\hat{M}_i,\;\,i=1,2,4$, is the decoded $M_i$ at receiver 1, and $\check{M}_i,\;\,i=1,3,5$, is the decoded $M_i$ at receiver 2.} 
	\vspace{-0pt}
	\label{Fig:DM-BC}
\end{figure}

\begin{figure*}[t]
		\begin{align}
			R_1+R_2+R_4&<I(U_0,U_1;Y_1)\label{eq:marton1},\\
			R_1+R_3+R_5&<I(U_0,U_2;Y_2)\label{eq:marton2},\\
			R_1+R_2+R_3+R_4&<I(U_0,U_1;Y_1)+I(U_2;Y_2\mid U_0)-I(U_1;U_2\mid U_0)\label{eq:marton3},\\
			R_1+R_2+R_3+R_5&<I(U_0,U_2;Y_2)+I(U_1;Y_1\mid U_0)-I(U_1;U_2\mid U_0)\label{eq:marton4},\\
			2R_1+R_2+R_3+R_4+R_5&<I(U_0,U_1;Y_1)+I(U_0,U_2;Y_2)-I(U_1;U_2\mid U_0)\label{eq:marton5}.
		\end{align}
	\vskip-7pt
	\hrulefill
	\vskip-15pt
\end{figure*}

\subsection{Contributions}
We consider the message setup for two-receiver broadcast channels that includes all possible message requests and side information configurations as special cases, i.e., each receiver (i) has both common and private-message requests, and (ii) knows part of the private message requested by the other receiver as side information. We propose a transmission scheme and derive an inner bound for the two-receiver memoryless broadcast channel. We show that this inner bound (i) establishes the capacity regions for two new classes, namely the deterministic channel and the more capable channel, and (ii) is tight for all classes of two-receiver broadcast channels with known capacity regions. A summary of the results is illustrated in Fig.~\ref{Fig:ResultSummary}.

\section{System Model}\label{Section:SystemModel}
We consider the two-receiver discrete-time memoryless broadcast channel $p(y_1,y_2\hskip-4.5pt\mid\hskip-4.5pt x)$, depicted in Fig. \ref{Fig:DM-BC},  with input $X\in\mathcal{X}$, and outputs $Y_1\in\mathcal{Y}_1$ and $Y_2\in\mathcal{Y}_2$. In this channel, $X^{n}=\left(X_1,X_2,\ldots,X_n\right)$ is the transmitted codeword, and $Y_{i}^{n}=\left(Y_{i,1},Y_{i,2},\ldots,Y_{i,n}\right),\;\,i=1,2$, is the channel-output sequence at receiver $i$.  The transmitted codeword is a function of source messages, $\mathbf{M}=\{M_i\}_{i=1}^5$. The source messages are independent, and $M_i$ is uniformly distributed over the set $\mathcal{M}_i=\{1,2,\ldots,2^{nR_i}\}$, i.e., transmitted at rate $R_i$ bits per channel use.

We define two sets corresponding to each receiver. $\mathbf{W}_1=\{M_1,M_2,M_4\}$ and $\mathbf{W}_2=\{M_1,M_3,M_5\}$ are the set of messages requested by receivers 1 and 2 respectively. $\mathbf{K}_1=\{M_5\}$ and $\mathbf{K}_2=\{M_4\}$ are the set of messages known a priori to receivers 1 and 2 respectively. For receiver 1, $M_2$ is the part of the private-message request which is not known a priori to the other receiver, and $M_4$ is the part which is known. For receiver 2, these are $M_3$ and $M_5$ respectively.

A $\left(2^{nR_1},2^{nR_2},2^{nR_3},2^{nR_4},2^{nR_5},n\right)$ code for the channel consists of an encoding function
\begin{align*}
f:\mathcal{M}_1\times \mathcal{M}_2 \times \mathcal{M}_3 \times \mathcal{M}_4 \times \mathcal{M}_5 \rightarrow \mathcal{X}^{n},
\end{align*}
where $\times$ denotes the Cartesian product, and $\mathcal{X}^{n}$ denotes the $n$-fold Cartesian product of $\mathcal{X}$. It also consists of decoding functions
\vskip-20pt
\begin{align*}
g_1:\mathcal{Y}_1^{n}\times \mathcal{M}_5 \rightarrow \mathcal{M}_1\times \mathcal{M}_2 \times \mathcal{M}_4,\\
g_2:\mathcal{Y}_2^{n}\times \mathcal{M}_4 \rightarrow \mathcal{M}_1\times \mathcal{M}_3 \times \mathcal{M}_5.
\end{align*}
Average probability of error for this code is defined as
\begin{multline*}
P_e^{(n)}=P((\hat{M}_1,\hat{M}_2,\hat{M}_4)\neq(M_1,M_2,M_4)\;\text{or} \\ \; (\check{M}_1,\check{M}_3,\check{M}_5)\neq(M_1,M_3,M_5)),
\end{multline*}
where $\hat{M}_i,\;\,i=1,2,4$, is the decoded $M_i$ at receiver 1, and $\check{M}_i,\;\,i=1,3,5$, is the decoded $M_i$ at receiver 2.
\begin{definition}
	A rate tuple $(R_1,R_2,R_3,R_4,R_5)$ is said to be \textit{achievable} for the channel if there exists a sequence of $\left(2^{nR_1},2^{nR_2},2^{nR_3},2^{nR_4},2^{nR_5},n\right)$ codes with $P_e^{(n)}\rightarrow 0$ as $n \rightarrow \infty$.
\end{definition}
\begin{definition}
	The \textit{capacity region} of the channel is the closure of the set of all achievable rate tuples $(R_1,R_2,R_3,R_4,R_5)$.
\end{definition}
\begin{definition}
A two-receiver memoryless broadcast channel is said to be deterministic if the channel outputs are deterministic functions of the channel input, i.e., $Y_i=\phi_i(X),\;\,i=1,2$. 
\end{definition}
\begin{definition}
A two-receiver memoryless broadcast channel is said to be more capable if $I\left(X;Y_1\right)\geq I\left(X;Y_2\right)$ for all input distributions~$p(x)$. 
\end{definition}

\section{Proposed Scheme and Inner Bound}\label{Section:Marton}
In this section, we propose a transmission scheme and derive an inner bound for the two-receiver memoryless broadcast channel with receiver message side information, stated as Theorem~\ref{theorem:marton}. The transmission scheme is constructed using Marton's coding scheme~\cite[p. 205]{NITBook}, superposition coding~\cite{BC}, and rate splitting.

\begin{theorem}\label{theorem:marton} A rate tuple $(R_1,R_2,R_3,R_4,R_5)$ for the two-receiver memoryless broadcast channel $p(y_1,y_2\mid x)$ is achievable if it satisfies \eqref{eq:marton1}--\eqref{eq:marton5} for some $p(u_0,u_1,u_2)$ and some function $x=\gamma(u_0,u_1,u_2)$.
\end{theorem}

\begin{figure*}[t]
		\begin{align}
		&\mathcal{E}_0:\left(U_0^n\left(1,1,1,1,1\right),U_1^n\left(1,1,1,1,1,1,l_1\right),U_2^n\left(1,1,1,1,1,1,l_2\right)\right)\notin\mathcal{T}_{\epsilon'}^n\text{ for all }l_1\text{ and }l_2,\label{eq:event0}\\
		&\mathcal{E}_{11}:(U_0^n\left(1,1,1,1,1\right),U_1^n(1,1,1,1,1,1,2^{nR'_1}),Y_1^n)\notin\mathcal{T}_{\epsilon_1}^n,\label{eq:event1}\\
		&\mathcal{E}_{12}:\left(U_0^n\left(1,1,1,1,1\right),U_1^n\left(1,1,1,1,1,m_{22},l_1\right),Y_1^n\right)\in\mathcal{T}_{\epsilon_1}^n\text{ for some }m_{22}\neq1\text{ and }l_1,\label{eq:event2}\\
		&\mathcal{E}_{13}:\left(U_0^n\left(m_1,m_4,1,m_{21},m_{31}\right),U_1^n\left(m_1,m_4,1,m_{21},m_{31},m_{22},l_1\right),Y_1^n\right)\in\mathcal{T}_{\epsilon_1}^n\label{eq:event3}\\
		&\hskip28em\text{ for some }m_{1}\neq1,m_4,m_{21},m_{31},m_{22} \text{ and }l_1.\nonumber
		\end{align}
	\hrulefill
\end{figure*}

\begin{IEEEproof}
	(\textit{Codebook Construction}) The codebook of the transmission scheme is formed from three subcodebooks which are constructed according to the distribution $p(u_0,u_1,u_2)$. Before subcodebook construction, using rate splitting, $M_i,\;\,i=2,3,$ is divided into two independent messages $M_{i1}$ at rate $R_{i1}$, and $M_{i2}$ at rate $R_{i2}$ such that $R_i=R_{i1}+R_{i2}$.
	
	The first subcodebook consists of i.i.d. codewords $u_0^n(m_1,m_4,m_5,m_{21},m_{31})$
	generated according to $\prod_{j=1}^{n}p_{U_0}(u_{0,j})$ for each $(m_1,m_4,m_5,m_{21},m_{31})$.
	
	The second subcodebook consists of codewords
	$u_1^n(m_1,m_4,m_5,m_{21},m_{31},m_{22},l_1)$
	generated according to $\prod_{j=1}^{n}p_{U_1\hskip-1pt\mid\hskip-1pt U_0}(u_{1,j}\mid u_{0,j}(m_1,m_4,m_5,m_{21},m_{31}))$~where $l_1\in\{1,2,\ldots,2^{nR'_1}\}$, i.e., for each $(m_1,m_4,m_5,m_{21},m_{31},m_{22})$, $2^{nR'_1}$ codewords are generated.
	
	The third subcodebook consists of codewords $u_2^n(m_1,m_4,m_5,m_{21},m_{31},m_{32},l_2)$ generated according to $\prod_{j=1}^{n}p_{U_2\hskip-1pt\mid\hskip-1ptU_0}(u_{2,j}\mid u_{0,j}(m_1,m_4,m_5,m_{21},m_{31}))$~where $l_2\in\{1,2,\ldots,2^{nR'_2}\}$, i.e., for each $(m_1,m_4,m_5,m_{21},m_{31},m_{32})$, $2^{nR'_2}$ codewords are generated. 
	
	(\textit{Encoding}) For the encoding, given $\{m_i\}_{i=1}^5$, we first find a pair $(l_1,l_2)$ such that 
	\begin{align*}
	\left(U_0^n\left(\cdot\right),U_1^n\left(\cdot,l_1\right),U_2^n\left(\cdot,l_2\right)\right)\in\mathcal{T}_{\epsilon'}^n,
	\end{align*}
	where $\mathcal{T}_{\epsilon'}^n$ is the set of jointly $\epsilon'$-typical $n$-sequences with respect to the considered distribution~\cite[p. 29]{NITBook}. If there is more than one pair, we arbitrary choose one of them, and if there does not exist one pair, we choose $(l_1,l_2)=(1,1)$. We then construct the transmitted codeword as $x_j=\gamma(u_{0,j}\left(\cdot\right),u_{1,j}\left(\cdot\right),u_{2,j}\left(\cdot\right)),\;\,j=1,2,\ldots,n$.
	
	(\textit{Decoding}) Receiver~1 decodes $\left(\hat{m}_1,\hat{m}_{21},\hat{m}_{22},\hat{m}_4\right)$, if it is the unique tuple for which we have 
	\begin{align*}
	\left(U_0^n\left(\cdot\right),U_1^n\left(\cdot,l_1\right),Y_1^n \right)\in\mathcal{T}_{\epsilon_1}^n\;\,\text{ for some}\;\,m_{31}\,\text{and } l_1;
	\end{align*}
	otherwise the error is declared.
	
	Receiver~2 similarly decodes $\left(\check{m}_1,\check{m}_{31},\check{m}_{32},\check{m}_5\right)$, if it is the unique tuple for which we have 
	\begin{align*}
	\left(U_0^n\left(\cdot\right),U_2^n\left(\cdot,l_2\right),Y_2^n\right)\in\mathcal{T}_{\epsilon_2}^n\;\,\text{ for some}\;\,m_{21}\,\text{and }l_2;
	\end{align*}
	otherwise the error is declared.
	
	We assume the transmitted messages are equal to one by the symmetry of code construction, and without loss of generality $(l_1,l_2)=(2^{nR'_1},2^{nR'_2})$. Hence, the error events at receiver~1 are \eqref{eq:event0}--\eqref{eq:event3}; note that there exist some other error events, but they yield redundant achievability conditions. The error events at receiver~2 are similarly written. Based on the error events, packing lemma~\cite[p. 45]{NITBook}, and mutual covering lemma~\cite[p. 208]{NITBook}, the achievability conditions are
	\begin{align*}
	R'_1+R'_2&>I\left(U_1;U_2\mid U_0\right),\\
	R_{22}+R'_1&<I\left(U_1;Y_1\mid U_0\right),\\
	R_1+R_{21}+R_{31}+R_4+R_{22}+R'_1&<I\left(U_0,U_1;Y_1\right),\\
	R_{32}+R'_2&<I\left(U_2;Y_2\mid U_0\right),\\
	R_1+R_{21}+R_{31}+R_5+R_{32}+R'_2&<I\left(U_0,U_2;Y_2\right).
	\end{align*}
	We finally perform Fourier-Motzkin elimination to obtain the region in \eqref{eq:marton1}--\eqref{eq:marton5}.
\end{IEEEproof}

\section {Capacity Results}
In this section, using the derived inner bound in Theorem~\ref{theorem:marton}, we establish the capacity regions for two new classes, i.e., the deterministic channel, stated as Theorem~\ref{theorem:deterministic}, and the more capable channel, stated as Theorem~\ref{theorem:morecapable}. We also show that our inner bound is tight for all classes of two-receiver broadcast channels whose capacity regions were known prior to this work.


\begin{theorem}\label{theorem:deterministic}
	The capacity region of the two-receiver deterministic broadcast channel with receiver message side information is the closure of the set of all rate tuples $(R_1,R_2,R_3,R_4,R_5)$, each satisfying
	\begin{align}
	R_1+R_2+R_4&\hskip-3pt<H(Y_1),\label{eq:detcond1}\\
	R_1+R_3+R_5&\hskip-3pt<H(Y_2),\label{eq:detcond2}\\
	R_1+R_2+R_3+R_4&\hskip-3pt<H(Y_1)+H(Y_2\mid U,Y_1),\label{eq:detcond3}\\
	R_1+R_2+R_3+R_5&\hskip-3pt<H(Y_2)+H(Y_1\mid U,Y_2),\label{eq:detcond4}\\
	2R_1+R_2+R_3+R_4+R_5&\hskip-3pt<\label{eq:detcond5}\\
	&\hskip-30ptI(U;Y_1)+H(Y_2)+H(Y_1\mid U,Y_2),\nonumber
	\end{align}
	for some $p(u,x)$.
\end{theorem}

We present the achievability proof in the following, and the converse proof in Appendix A.
\begin{IEEEproof}
	\textit{(Achievability)} The achievability part of Theorem~\ref{theorem:deterministic} is proved by setting $(U_0,U_1,U_2)=(U,Y_1,Y_2)$ in \eqref{eq:marton1}--\eqref{eq:marton5}.   
\end{IEEEproof}


\begin{theorem}\label{theorem:morecapable}
	The capacity region of the two-receiver more capable broadcast channel with receiver message side information is the closure of the set of all rate tuples $(R_1,R_2,R_3,R_4,R_5)$, each satisfying
	\begin{align}
	R_1+R_3+R_5&<I(U;Y_2),\label{eq:morecapcond1}\\
	R_1+R_2+R_3+R_5&<I(U;Y_2)+I(X;Y_1\mid U),\label{eq:morecapcond2}\\
	R_1+R_2+R_3+R_4&<I(X;Y_1),\label{eq:morecapcond3}
	\end{align}
	for some $p(u,x)$.
\end{theorem}

We present the achievability proof in the following, and the converse proof in Appendix B.
\begin{IEEEproof}
	\textit{(Achievability)} The achievability of Theorem~\ref{theorem:morecapable} is proved by setting $(U_0,U_1,U_2)=(U,X,0)$ in \eqref{eq:marton1}--\eqref{eq:marton5}. Note that $U_2=0$ implies that $M_{31}=M_3$, and $R'_1=0$. 
\end{IEEEproof}

\subsection{Discussion on Prior Known Results}
In this subsection, we show that the derived inner bound in Theorem~\ref{theorem:marton} is tight for all classes of two-receiver broadcast channels with known capacity regions, as depicted in Fig.~\ref{Fig:ResultSummary}.

The capacity region of the two-receiver memoryless broadcast channel with complementary side information is achieved by multiplexing all the requested messages in only one codebook \cite{SWoverBC,BCwithSI2UsersOechtering}. This scheme is a special case our scheme obtained by setting $(U_0,U_1,U_2)=(X,0,0)$. Note that $M_2$ and $M_3$ are equal to zero in this message setup. 

The capacity region of the two-receiver memoryless broadcast channel with degraded message sets is achieved by superposition coding~\cite{BCwithSI2UsersKramer}. This scheme is a special case of our scheme obtained by setting $(U_0,U_1,U_2)=(U,X,0)$ or $(U_0,U_1,U_2)=(U,0,X)$ depending on whether receiver~1 or receiver~2 needs to decode the whole set of the source messages, respectively. Note that either $M_2$ or $M_3$ is equal to zero in this message setup.

The AWGN broadcast channel and the less noisy broadcast channel are a subset of the more capable broadcast channel~\cite{NITBook}, as depicted in Fig.~\ref{Fig:ResultSummary}, then our scheme can also achieve their capacity regions.

\begin{figure*}[t]
	\setcounter{equation}{22}
	\begin{align}
	n(R_1+R_2+R_4)\hskip-2pt\leq& I\left(M_1,M_2,M_4;Y_1^n\mid M_5\right)+n\epsilon_n,\label{eq:detconverse1}\\
	n(R_1+R_3+R_5)\hskip-2pt\leq& I\left(M_1,M_3,M_5;Y_2^n\mid M_4\right)+n\epsilon_n,\label{eq:detconverse2}\\
	n(R_1+R_2+R_3+R_4)\hskip-2pt\leq& I\left(M_1,M_4;Y_1^n\mid M_5\right)+I\left(M_2,M_3;Y_1^n,Y_2^n\mid M_1,M_4,M_5\right)+3n\epsilon_n,\label{eq:detconverse3}\\
	n(R_1+R_2+R_3+R_5)\hskip-2pt\leq& I\left(M_1,M_5;Y_2^n\mid M_4\right)+I\left(M_2,M_3;Y_1^n,Y_2^n\mid M_1,M_4,M_5\right)+3n\epsilon_n,\label{eq:detconverse4}\\
	\hskip-6.5pt n(2R_1\hskip-2pt+\hskip-2ptR_2\hskip-2pt+\hskip-2ptR_3+R_4+R_5)\hskip-2pt\leq& I\left(M_1,M_4;Y_1^n\mid M_5\right)\hskip-2pt+\hskip-2ptI\left(M_1,M_5;Y_2^n\mid M_4\right)\hskip-2pt+\hskip-2ptI\left(M_2,M_3;Y_1^n,Y_2^n\mid M_1,M_4,M_5\right)\hskip-2pt+\hskip-2pt4n\epsilon_n\label{eq:detconverse5}.
	\end{align}
	\hrulefill
\end{figure*}

\begin{figure*}[b]
    \vskip-5pt
    \hrulefill
	\setcounter{equation}{28}
	\begin{align}
	I\left(M_1,M_5;Y_2^n\mid M_4\right)&+I\left(M_2,M_3;Y_1^n,Y_2^n\mid M_1,M_4,M_5\right)\nonumber\\
	&=\underset{\text{part 1}}{\underbrace{I\left(M_1,M_5;Y_2^n\mid M_4\right)+I\left(M_2,M_3;Y_1^n\mid M_1,M_4,M_5\right)}}+\underset{\text{part 2}}{\underbrace{I\left(M_2,M_3;Y_2^n\mid M_1,M_4,M_5,Y_1^n\right)}}\label{eq:detconverse41}.
	\end{align}

	\begin{equation}
	I\left(M_1,M_5;Y_2^n\mid M_4\right)+I\left(M_2,M_3;Y_1^n\mid M_1,M_4,M_5\right)\hskip260pt\label{eq:detconverse42}
	\end{equation}
	
	$\hskip80pt=\sum_{i=1}^{n}I\left(M_1,M_5;Y_{2,i}\mid M_4,Y_{2,i+1}^{n}\right)+\sum_{i=1}^{n}I\left(M_2,M_3;Y_{1,i}\mid M_1,M_4,M_5,Y_1^{i-1}\right)$
	
	$\hskip80pt\leq\sum_{i=1}^{n}I\left(M_1,M_4,M_5,Y_{2,i+1}^{n};Y_{2,i}\right)+\sum_{i=1}^{n}I\left(M_2,M_3,Y_{2,i+1}^{n};Y_{1,i}\mid M_1,M_4,M_5,Y_1^{i-1}\right)$
	
	$\hskip80pt=\sum_{i=1}^{n}I\left(M_1,M_4,M_5,Y_1^{i-1},Y_{2,i+1}^{n};Y_{2,i}\right)-\sum_{i=1}^{n}I\left(Y_1^{i-1};Y_{2,i}\mid M_1,M_4,M_5,Y_{2,i+1}^{n} \right)$
	
	$\hskip95pt+\sum_{i=1}^{n}I\left(Y_{2,i+1}^{n};Y_{1,i}\mid M_1,M_4,M_5,Y_1^{i-1}\right)+\sum_{i=1}^{n}I\left(M_2,M_3;Y_{1,i}\mid M_1,M_4,M_5,Y_1^{i-1},Y_{2,i+1}^{n}\right)$
	
	$\hskip79pt\overset{(a)}{=}\sum_{i=1}^{n}I\left(M_1,M_4,M_5,Y_1^{i-1},Y_{2,i+1}^{n};Y_{2,i}\right)+\sum_{i=1}^{n}I\left(M_2,M_3;Y_{1,i}\mid M_1,M_4,M_5,Y_1^{i-1},Y_{2,i+1}^{n}\right)$
	
	$\hskip80pt=\sum_{i=1}^{n}I\left(U_i;Y_{2,i}\right)+\sum_{i=1}^{n}H\left(Y_{1,i}\mid U_i\right).$
	
	\begin{equation}
	I\left(M_2,M_3;Y_2^n\mid M_1,M_4,M_5,Y_1^n\right)\hskip340pt\label{eq:detconverse43}
	\end{equation}
	$\hskip110pt=H\left(Y_2^n\mid M_1,M_4,M_5,Y_1^{n}\right)=\sum_{i=1}^{n}H\left(Y_{2,i}\mid M_1,M_4,M_5,Y_{2,i+1}^{n},Y_1^{n}\right)$
	
	$\hskip110pt\leq\sum_{i=1}^{n}H\left(Y_{2,i}\mid M_1,M_4,M_5,Y_{2,i+1}^{n},Y_{1,i},Y_1^{i-1}\right)=\sum_{i=1}^{n}H\left(Y_{2,i}\mid U_i,Y_{1,i}\right).$

\end{figure*}

\section{Conclusion}
We proposed a transmission scheme and derived an inner bound for the two-receiver memoryless broadcast channel with receiver message side information. We considered the general message setup which includes all possible message requests and side information configurations as special cases. Our proposed scheme is a unified capacity-achieving scheme for all classes of two-receiver broadcast channels whose capacity regions had been previously established, and for two new classes, i.e., the deterministic channel and the more capable channel.


\section*{Appendix A}\label{Appendix:deterministic}
In this section, we present the converse proof for the two-receiver deterministic broadcast channel with receiver message side information. The proof is based on the converse proof for the channel without receiver message side information~\cite{DeterminsticBCwithCommon}.
\begin{IEEEproof}\textit{(Converse Proof)}
By Fano's inequality \cite[p. 19]{NITBook}, we have
\setcounter{equation}{17}
\begin{align}
H\left(M_1,M_2,M_4\mid Y_1^n,M_5\right)\leq n\epsilon_{1,n}\label{eq:fano1},\\
H\left(M_1,M_3,M_5\mid Y_2^n,M_4\right)\leq n\epsilon_{2,n}, \label{eq:fano2}
\end{align}
where $\epsilon_{i,n}\rightarrow 0$ as $n\rightarrow \infty$ for $i=1,2$. For the sake of simplicity, we use $\epsilon_{n}$ instead of $\epsilon_{i,n}$ for the remainder. The inequalities in \eqref{eq:fano1}--\eqref{eq:fano2} also lead to the following inequalities,
\begin{align}
H\left(M_1,M_4\mid Y_1^n,M_5\right)\leq n\epsilon_{n},\label{eq:fano3}\\
H\left(M_1,M_5\mid Y_2^n,M_4\right)\leq n\epsilon_{n},\label{eq:fano4}\\
H\left(M_2,M_3\mid Y_1^n,Y_2^n,M_1,M_4,M_5\right)\leq 2n\epsilon_n.\label{eq:fano5}
\end{align}

Using \eqref{eq:fano1}--\eqref{eq:fano5}, if a rate tuple $(R_1,R_2,R_3,R_4,R_5)$ is achievable, then it must satisfy \eqref{eq:detconverse1}--\eqref{eq:detconverse5}. 

Inequalities \eqref{eq:detconverse1}--\eqref{eq:detconverse5} yield conditions \eqref{eq:detcond1}--\eqref{eq:detcond5} respectively. To this end, we use the Csisz\'{a}r sum identity \cite[p. 25]{NITBook} based on which we have
\setcounter{equation}{27}
\begin{equation}\label{csiszar}
\begin{split}
\sum_{i=1}^{n}I(&Y_1^{i-1};Y_{2,i}\mid M_1,M_4,M_5,Y_{2,i+1}^n)\\
&=\sum_{i=1}^{n}I\left(Y_{2,i+1}^n;Y_{1,i}\mid M_1,M_4,M_5,Y_1^{i-1}\right),
\end{split}
\end{equation} 
where $Y_1^{i-1}=\left(Y_{1,1},Y_{1,2},\ldots,Y_{1,i-1}\right)$ and $Y_{2,i+1}^{n}=\left(Y_{2,i+1},Y_{2,i+2},\ldots,Y_{2,n}\right)$. 
We also need to define the auxiliary random variable as
\begin{align*}
	U_i=\left(M_1,M_4,M_5,Y_1^{i-1},Y_{2,i+1}^n\right).
\end{align*}
We here show only how inequalities \eqref{eq:detconverse1} and \eqref{eq:detconverse4} yield conditions \eqref{eq:detcond1} and \eqref{eq:detcond4} respectively. We just need to follow similar steps for \eqref{eq:detconverse2}, \eqref{eq:detconverse3}, and \eqref{eq:detconverse5}.

In \eqref{eq:detconverse1}, we expand the mutual information term as follows
\begin{align*}
I(M_1,M_2, & M_4;Y_1^n \mid M_5)\\
&\leq I\left(M_1,M_2,M_4;Y_1^n\mid M_3,M_5\right)\\
&=H\left(Y_1^n\mid M_3,M_5\right)\leq H\left(Y_1^n\right)\leq\sum_{i=1}^n H\left(Y_{1,i}\right).
\end{align*}
Then, since $\epsilon_n\rightarrow 0$ as $n\rightarrow \infty$, by using the standard time-sharing argument \cite[p. 114]{NITBook}, we have
\vspace{0pt}
\begin{align*}
R_1+R_2+R_4\leq H(Y_1).
\end{align*}

In~\eqref{eq:detconverse4}, we first expand the mutual information terms as in \eqref{eq:detconverse41}. We then expand part 1 of \eqref{eq:detconverse41} as in \eqref{eq:detconverse42} where $(a)$ follows from \eqref{csiszar}. We also expand part 2 of \eqref{eq:detconverse41} as in \eqref{eq:detconverse43}. Finally, since $\epsilon_n\rightarrow 0$ as $n\rightarrow \infty$, and
\begin{align*}
I\left(U_i;Y_{2,i}\right)& + H\left(Y_{1,i}\mid U_i\right)+H\left(Y_{2,i}\mid U_i,Y_{1,i}\right)\\
& =H\left(Y_{2,i}\right)+H\left(Y_{1,i}\mid U_i\right)-I\left(Y_{1,i};Y_{2,i}\mid U_i\right)\\
& =H\left(Y_{2,i}\right)+H\left(Y_{1,i}\mid U_i,Y_{2,i}\right),
\end{align*}
by using the standard time-sharing argument, we have
\begin{align*}
R_1+R_2+R_3+R_5\leq H(Y_2)+H(Y_1\mid U,Y_2).
\end{align*}
\end{IEEEproof}

\begin{figure*}[t]
	\setcounter{equation}{31}
	\begin{align}
	n(R_1+R_3+R_5)\hskip-2pt\leq& I\left(M_1,M_3,M_5;Y_2^n\mid M_4\right)+n\epsilon_n,\label{eq:mocapconverse1}\\
	n(R_1+R_2+R_3+R_5)\hskip-2pt\leq& I\left(M_2;Y_1^n\mid M_1,M_3,M_4,M_5\right)+I\left(M_1,M_3,M_5;Y_2^n\mid M_4\right)+2n\epsilon_n,\label{eq:mocapconverse2}\\
	n(R_1+R_2+R_3+R_4)\hskip-2pt\leq& I\left(M_1,M_2,M_4;Y_1^n\mid M_5\right)+I\left(M_3;Y_2^n\mid M_1,M_2,M_4,M_5\right)+2n\epsilon_n.\label{eq:mocapconverse3}
	\end{align}
	\hskip-15pt
	\hrulefill
	\hskip-15pt
\end{figure*}

\begin{figure*}[b]
    \hskip-15pt
    \hrulefill
    \hskip-15pt
	\setcounter{equation}{36}
\begin{align}
I\left(M_2;Y_1^n\mid M_1,M_3,M_4,M_5\right)+I\left(M_1,M_3,M_5;Y_2^n\mid M_4\right)\hskip240pt\label{eq:morecabproof21}
\end{align}
$\hskip90pt=\sum_{i=1}^{n}I\left(M_2;Y_{1,i}\mid M_1,M_3,M_4,M_5,Y_1^{i-1}\right)+\sum_{i=1}^{n}I\left(M_1,M_3,M_5;Y_{2,i}\mid M_4,Y_{2,i+1}^n\right)$

$\hskip90pt\leq\sum_{i=1}^{n}I\left(M_2,Y_{2,i+1}^n;Y_{1,i}\mid M_1,M_3,M_4,M_5,Y_1^{i-1}\right)+\sum_{i=1}^{n}I\left(M_1,M_3,M_4,M_5,Y_{2,i+1}^n;Y_{2,i}\right)$

%

$\hskip90pt\overset{(a)}{=}\sum_{i=1}^{n}I\left(M_2;Y_{1,i}\mid M_1,M_3,M_4,M_5,Y_1^{i-1},Y_{2,i+1}^n\right)+\sum_{i=1}^{n}I\left(M_1,M_3,M_4,M_5,Y_1^{i-1},Y_{2,i+1}^n;Y_{2,i}\right)$

$\hskip90pt\overset{(b)}{=}\sum_{i=1}^{n}I\left(M_2;Y_{1,i}\mid U_i\right)+\sum_{i=1}^{n}I\left(U_i;Y_{2,i}\right)$

$\hskip90pt\leq\sum_{i=1}^{n}I\left(M_2,X_i;Y_{1,i}\mid U_i\right)+\sum_{i=1}^{n}I\left(U_i;Y_{2,i}\right)=\sum_{i=1}^{n}I\left(X_i;Y_{1,i}\mid U_i\right)+\sum_{i=1}^{n}I\left(U_i;Y_{2,i}\right).$
\end{figure*}

\section*{Appendix B}\label{Appendix:morecapable}
In this section, we present the converse proof for the two-receiver more capable broadcast channel with receiver message side information. The proof is based on the converse proof for the channel without receiver message side information~\cite{MoreCapableBCwithCommon}.

\begin{IEEEproof} \textit{(Converse Proof)}
Using \eqref{eq:fano1} and \eqref{eq:fano2}, if a rate tuple $(R_1,R_2,R_3,R_4,R_5)$ is achievable, then it must satisfy \eqref{eq:mocapconverse1}--\eqref{eq:mocapconverse3}. 

In \eqref{eq:mocapconverse1}, we expand the mutual information term as follows

$I(M_1,M_3,M_5;Y_2^n\mid M_4)$

$\hskip40pt=\sum_{i=1}^{n}I(M_1,M_3,M_5;Y_{2,i}\mid M_4,Y_{2,i+1}^n)$

$\hskip40pt\leq\sum_{i=1}^{n}I(M_1,M_3,M_4,M_5,Y_{2,i+1}^n;Y_{2,i})$

$\hskip40pt\leq\sum_{i=1}^{n}I(M_1,M_3,M_4,M_5,Y_{1}^{i-1},Y_{2,i+1}^n;Y_{2,i}).$\\
This results in
\setcounter{equation}{34}
\begin{align}
n(R_1+R_3+R_5)&\leq \sum_{i=1}^{n}I(U_i;Y_{2,i})+n\epsilon_n,\label{eq:morecabproof1}
\end{align}
where the auxiliary random variable $U_i$ is defined as
\begin{align}
U_i=\left(M_1,M_3,M_4,M_5,Y_1^{i-1},Y_{2,i+1}^n\right).\label{auxiliarymorecap}
\end{align}

In \eqref{eq:mocapconverse2}, we expand the mutual information terms as in \eqref{eq:morecabproof21} where $(a)$ follows from the Csisz\'{a}r sum identity~\cite[p. 25]{NITBook} and $(b)$ from \eqref{auxiliarymorecap}. This results in
\setcounter{equation}{37}
\begin{multline}
n(R_1+R_2+R_3+R_5)\leq\\ \sum_{i=1}^{n}\left(I(U_i;Y_{2,i})+I(X_i;Y_{1,i}\mid U_i)\right)+2n\epsilon_n,\label{eq:morecabproof22}
\end{multline}

In \eqref{eq:mocapconverse3}, by expanding the mutual information terms similar to the ones in \eqref{eq:mocapconverse2}, we have
\begin{multline}\label{eq:morecabproof31}
n(R_1+R_2+R_3+R_4)\leq\\ \sum_{i=1}^{n}\left(I(V_i;Y_{1,i})+I(X_i;Y_{2,i}\mid V_i)\right)+2n\epsilon_n,
\end{multline}
where $V_i$ is defined as
\begin{align*}
V_i=\left(M_1,M_2,M_4,M_5,Y_1^{i-1},Y_{2,i+1}^n\right).
\end{align*}
Inequality \eqref{eq:morecabproof31} yields   
\begin{align}\label{eq:morecabproof32}
n(R_1+R_2+R_3+R_4)&\leq \sum_{i=1}^{n}I(X_i;Y_{1,i})+2n\epsilon_n,
\end{align}
because for the more capable channel, if $V\rightarrow X\rightarrow (Y_1,Y_2)$ form a Markov chain, then we have~\cite[p. 123]{NITBook}
\begin{align*}
I(X;Y_2\mid V)\leq I(X;Y_1\mid V).
\end{align*}

Since $\epsilon_n\rightarrow0$ as $n\rightarrow\infty$, by using the standard time-sharing argument~\cite[p. 114]{NITBook} for \eqref{eq:morecabproof1}, \eqref{eq:morecabproof22} and \eqref{eq:morecabproof32}, the converse proof is complete.
\end{IEEEproof}
\bibliographystyle{IEEEtran}

\begin{thebibliography}{10}
	\providecommand{\url}[1]{#1}
	\csname url@samestyle\endcsname
	\providecommand{\newblock}{\relax}
	\providecommand{\bibinfo}[2]{#2}
	\providecommand{\BIBentrySTDinterwordspacing}{\spaceskip=0pt\relax}
	\providecommand{\BIBentryALTinterwordstretchfactor}{4}
	\providecommand{\BIBentryALTinterwordspacing}{\spaceskip=\fontdimen2\font plus
		\BIBentryALTinterwordstretchfactor\fontdimen3\font minus
		\fontdimen4\font\relax}
	\providecommand{\BIBforeignlanguage}[2]{{%
			\expandafter\ifx\csname l@#1\endcsname\relax
			\typeout{** WARNING: IEEEtran.bst: No hyphenation pattern has been}%
			\typeout{** loaded for the language `#1'. Using the pattern for}%
			\typeout{** the default language instead.}%
			\else
			\language=\csname l@#1\endcsname
			\fi
			#2}}
	\providecommand{\BIBdecl}{\relax}
	\BIBdecl
	
	\bibitem{BC}
	T.~M. Cover, ``Broadcast channels,'' \emph{{IEEE} Trans. Inf. Theory}, vol.~18,
	no.~1, pp. 2--14, Jan. 1972.
	
	\bibitem{TWRC}
	Y.~Wu, P.~A. Chou, and S.~Y. Kung, ``Information exchange in wireless networks
	with network coding and physical-layer broadcast,'' in \emph{Proc. Conf. Inf.
		Sci. Syst. (CISS)}, Baltimore, MD, Mar. 2005.
	
	\bibitem{BCwithSI2UsersOechtering}
	T.~J. Oechtering, C.~Schnurr, I.~Bjelakovic, and H.~Boche, ``Broadcast capacity
	region of two-phase bidirectional relaying,'' \emph{{IEEE} Trans. Inf.
		Theory}, vol.~54, no.~1, pp. 454--458, Jan. 2008.
	
	\bibitem{SWoverBC}
	E.~Tuncel, ``Slepian-{W}olf coding over broadcast channels,'' \emph{{IEEE}
		Trans. Inf. Theory}, vol.~52, no.~4, pp. 1469--1482, Apr. 2006.
	
	\bibitem{BCwithSI2UsersKramer}
	G.~Kramer and S.~Shamai, ``Capacity for classes of broadcast channels with
	receiver side information,'' in \emph{Proc. IEEE Inf. Theory Workshop (ITW)},
	Lake Tahoe, CA, Sept. 2007, pp. 313--318.
	
	\bibitem{BCwithSI2UsersGeneral}
	Y.~Wu, ``Broadcasting when receivers know some messages a priori,'' in
	\emph{Proc. IEEE Int. Symp. Inf. Theory (ISIT)}, Nice, France, June 2007, pp.
	1141--1145.
	
	\bibitem{BCwithSI3UsersCommonMessage}
	T.~J. Oechtering, M.~Wigger, and R.~Timo, ``Broadcast capacity regions with
	three receivers and message cognition,'' in \emph{Proc. IEEE Int. Symp. Inf.
		Theory (ISIT)}, Cambridge, MA, July 2012, pp. 388--392.
	
	\bibitem{BCwithSI3UsersLinearDeterministic}
	J.~W. Yoo, T.~Liu, and F.~Xue, ``Broadcasting with receiver message side
	information : A deterministic approach,'' in \emph{Proc. 46th Annu. Allerton
		Conf. Commun. Control Comput.}, Monticello, IL, Sept. 2008, pp. 746--752.
	
	\bibitem{Capacity3UsersPrivateMessage}
	B.~Asadi, L.~Ong, and S.~J. Johnson, ``The capacity of three-receiver {AWGN}
	broadcast channels with receiver message side information,'' in \emph{Proc.
		IEEE Int. Symp. Inf. Theory (ISIT)}, Honolulu, HI, June/July 2014, pp.
	2899--2903.
	
	\bibitem{Group4andGroup7}
	------, ``Coding schemes for a class of receiver message side information in
	{AWGN} broadcast channels,'' in \emph{Proc. IEEE Inf. Theory Workshop (ITW)},
	Hobart, Australia, Nov. 2014, pp. 571--575.
	
	\bibitem{Capacity3UsersPrivateMessageParallel}
	J.~Sima and W.~Chen, ``Joint network and {G}elfand-{P}insker coding for
	3-receiver {G}aussian broadcast channels with receiver message side
	information,'' in \emph{Proc. IEEE Int. Symp. Inf. Theory (ISIT)}, Honolulu,
	HI, June/July 2014, pp. 81--85 [Revised Version] Available:
	http://arxiv.org/abs/1407.8409v2.
	
	\bibitem{NITBook}
	A.~{El G}amal and Y.~H. Kim, \emph{Network Information Theory}.\hskip 1em plus
	0.5em minus 0.4em\relax Cambridge University Press, 2011.
	
	\bibitem{DeterminsticBCwithCommon}
	T.~S. Han, ``The capacity region for the deterministic broadcast channel with a
	common message,'' \emph{{IEEE} Trans. Inf. Theory}, vol.~27, no.~1, pp.
	122--125, Jan. 1981.
	
	\bibitem{MoreCapableBCwithCommon}
	A.~{El G}amal, ``The capacity of a class of broadcast channels,'' \emph{{IEEE}
		Trans. Inf. Theory}, vol.~25, no.~2, pp. 166--169, Mar. 1979.
	
\end{thebibliography}

\end{document}